\documentclass[12pt]{amsart}
\usepackage{a4wide}
\usepackage{amsmath}
\usepackage{amssymb}

\newtheorem{stw}{{\bf Proposition}}
\newtheorem{tw}{{\bf Theorem}}
\newtheorem{lem}{{\bf Lemma}}
\newtheorem{df}{{\bf Definition}}

\newenvironment{dow}{\noindent{\bf\em Proof.}\,}
{\hspace*{\fill}\(\boxtimes\)\newline}

\newcounter{liczprz}
\newenvironment{prz}{\refstepcounter{liczprz}{\noindent
\bf Example~\theliczprz.}}{\hspace*{\fill}\(\lozenge\)\newline}

\newcommand{\R}{{\mathbb R}}
\newcommand{\dach}[1]{{\widetilde{#1}}}
\newcommand{\sprz}[1]{{\overline{#1}}}
\newcommand{\impl}{{\Rightarrow}}
\newcommand{\dstrz}{{\Leftrightarrow}}
\newcommand{\Pb}{P^{\flat}}
\newcommand{\supp}{\operatorname{supp}}
\newcommand{\sign}{\operatorname{sign}}

\begin{document}

\title[Playing cooperatively]
{Playing cooperatively with possibly treacherous partner}

\author[Le\'{s}niak]{K. Le\'{s}niak}
\date{\today}




\email{much@mat.umk.pl, klesniak.much@gmail.com}

\begin{abstract}
We investigate an alternative concept of Nash equilibrium,
m-equilibrium, which slightly resembles Harsanyi-Selten risk
dominant equilibrium although it is a different notion.
M-equilibria provide nontrivial solutions of normal form games as
shown by comparison of the Prisoner's Dilemma with the Traveler's
Dilemma. They are also resistant on the deep iterated elimination
of dominated strategies.
\end{abstract}

\maketitle%

\section{Introduction}

The games of interest here are 
\emph{two person general sum games in normal form}. 
Concerning notation and definitions the reader is
asked to consult the next Section.

First one has to provide additional information about the game.
The main assumption, usually made implicitly, is that the players
can communicate with each other. This remedies the coordination
problem quoted below (see \cite{CommComplex} for complexity
issues).

\begin{prz}[Coordination]\label{Coordination}
Let \(S_1=S_2=\{1,2,3\}\),
\[G= \left[\begin{array}{ccc}
{[2,2]} & {[0,0]} & {[0,0]} \\
{[0,0]} & {[1,1]} & {[0,0]} \\
{[0,0]} & {[0,0]} & {[2,2]} \\
\end{array}\right].\]
Among three equilibria \((1,1)\), \((2,2)\) and \((3,3)\) only two
are pleasant (as Pareto dominant), namely \((1,1)\) and \((3,3)\).
If players cannot communicate, then they have to use randomization
(e.g., coin flipping). The Bernoulli scheme would let them
synchronize choices with small probability of failure (cf. Theorem 11.3
in \cite[chap.11.4]{Rendezvous}).
\end{prz}

Thus we see that some amount of communication (say pre-play/cheap talk) 
and a sort of cooperation cannot be dismissed even in the case of 
such competitive/``selfish" notion like the Nash equilibrium
(e.g., \cite{Aumann, CommCoord, NashNeedsCooper}). 
The classic stag hunt game exploits another
issue of miscoordination -- the Wald criterion of worst possible
scenario. In the vein of a stag hunt's strategic security consider

\begin{prz}\label{HighThreat}
Let \(S_1=S_2=\{1,2\}\),
\[G=\left[\begin{array}{cc}
{[4,4]} & {[1,4]} \\
{[4,1]} & {[3,3]} \\
\end{array}\right].\]
The pairs \((1,1)\) and \((2,2)\) are Nash equilibria with
\((1,1)\) being Pareto dominant. Nevertheless one cannot guarantee
that previously agreed among players equilibrium \((1,1)\) would
be realized in practice. If the player is confident in fair play
of his partner, he might switch strategy without any loss of
income. The pair of strategies \((2,2)\) is threat-safe although
yields smaller payoffs than \((1,1)\).
\end{prz}

Players make their final decisions independently of others.
Therefore communication provides only weak cooperation
(\cite{FundEquilibrium}). No one can force fair play, even if it
is profitable for all (free rider's problem).

Largely discussed traveler's dilemma (seen sometimes as the extension
of prisoner's dilemma) underlines anomalous behavior in widely
accepted procedure called an iterative elimination of dominated
strategies (\cite{Basu, BasuExperim, Spanish, Gintis, HalpernPass}). 
Unlike the original formulation our assumes communication between players.

\begin{prz}[Traveler's dilemma]\label{traveler}
Let \(S_1=S_2=\{2,3,\ldots,100\}\), 
\(P_1(x,y)=P_2(y,x)= \min(x,y)+ 2\cdot \sign(y-x)\) 
for \(x\in S_1\), \(y\in S_2\),
\(G=(S_1,S_2; P_1,P_2)\). Then \((2,2)\) is the only Nash
equilibrium of \(G\). It arises through the elimination of dominated
strategies, although most strategy pairs Pareto dominate it.

Observe that the players could choose a pair of strategies which
yields much higher payoffs than \((2,2)\). Moreover, the player
can still play very profitably after his partner betrayed and
switched strategy to get higher payoffs. If more than \(4\%\)
partners play ``moderately" (at least \(54\)), then we can expect
higher gain from playing ``dummy" \(100\) than from playing ``wise"
\(2\). If more than \(10\%\) partners play ``high" (at least
\(90\)), then playing \(100\) we can expect over \(400\%\) income
of that which we could earn playing \(2\).  
\end{prz}

Our point here is that one should calculate secure gains incorporating
possible threats from his partner. This allows sometimes for much
higher payoffs than those arising in the Nash equilibrium. 
That was the main motivation for introducing m-equilibria as we do
in Section~\ref{sec:m-equilibrium}.

We silently assume that the games under consideration are 
\emph{not repeated} and \emph{one-shot}; see also the discussion
around mixed equilibria in Examples \ref{hidecoin} and \ref{pennies}. 
We understand the payoff to be NTU (nontransferable utility); that some
``transfers" are still possible ensures us Example \ref{CournotDuopoly} 
in Section \ref{sec:motivexamples} and informal discussion of fair choice 
among equilibria in Section \ref{sec:equilibriumselection}.

For standard notions and theorems of game theory we refer to the
textbook \cite{Strategy} (comp. \cite{StratEquilibrium, AubinOpt}).
Throughout the paper the language of multivalued (or set-valued)
analysis shall be utilized in several places (consult 
\cite{HandbookMulti, Beer, AubinOpt}).

\section{Notation and definitions}

Let \(\Gamma= (\Sigma_1, \Sigma_2; P_1,P_2:\Sigma_1\times\Sigma_2\to\R)\)
be a two person game (in normal form).
\(\Sigma_i\) is the \textit{set of strategies} and
\(P_i\) is the \textit{payoff function} of the \(i\)-th player.

An accent will be put further on the case of a \textit{finite game} 
\(G=(S_1,S_2; W_1,W_2: S_1\times S_2 \to\R)\), i.e., the game with 
the finite strategy sets \(S_1,S_2\), and its \textit{mixed extension} 
\(\Delta(G)= (\Delta(S_1),\Delta(S_2); EW_1,EW_2: \Delta(S_1)\times\Delta(S_2)\to\R)\).
\(\Delta(S)\) stands for the standard simplex of probabilistic measures 
(\textit{mixed strategies}) spanned on the finite set \(S\) of (\textit{pure}) strategies;
\(S\subset\Delta(S)\) due to the identification via Dirac measures: 
\(S\ni x \mapsto \delta_{x}\in\Delta(S)\).
The \textit{expected payoffs} are given by
\[EW_i(\rho_1,\rho_2)=\sum_{(x,y)\in S_1\times S_2} 
{\rho}_1(x)\cdot W_i(x,y)\cdot {\rho}_2(y)\]
for \({\rho}_i\in\Delta(S_i)\), \(i=1,2\).
Finite games shall appear in the examples as bimatrices
\(\left[\begin{array}{c}
[W_1(x,y),W_2(x,y)]_{(x,y)\in S_1\times S_2}
\end{array}\right]\), 
\(S_1\ni x\) -- the number of the row, \(S_2\ni y\) -- the number of the column.

It is customary to employ the convention
\(((\sigma_{i},\sigma_{-i})) = (\sigma_1,\sigma_2) \in\Sigma_{1}\times\Sigma_{2}\),
\(\Sigma_{-i}=\Sigma_{3-i}\) and \(P_{-i}=P_{3-i}\) which
emphasizes the role of the \(i\)-th player (\(i=1,2\)); the same
applies to \(((x,y))\in {S}_{1}\times {S}_{2}\),
\(((\pi,y))\in\Delta(S_1)\times\Delta(S_2)\), \(W_{-i}\) etc.

For technical reasons we shall tacitly assume payoff functions \(P_i\)
to be bounded from below in the co-player's variable, i.e.,
\(\inf_{\sigma_{-i}\in\Sigma_{-i}}\, P_i((\sigma_i,\sigma_{-i})) >-\infty\) for 
\(\sigma_{i}\in\Sigma_{i}\), \(i=1,2\). The continuity of \(P_i\) is not assumed and,
in the case of finite strategy sets \(S_i\), of no use.

The game \(\Gamma\) is said to be \textit{strictly competitive}, if
\[\forall_{(\sigma_1,\sigma_2),(\sigma_1',\sigma_2')\in\Sigma_1\times\Sigma_2}\;\;
P_1(\sigma_1',\sigma_2')>P_1(\sigma_1,\sigma_2) \dstrz
P_2(\sigma_1',\sigma_2')<P_2(\sigma_1,\sigma_2).\]
Let us note that strictly competitive games are essentially zero-sum games 
(see \cite{Adler}).

We call \(\Gamma\) \textit{quantitatively symmetric}, if 
\(P_1(\sigma_1,\sigma_2)=P_2(\sigma_2,\sigma_1)\) for 
\(\sigma_1,\sigma_2\in\Sigma_1=\Sigma_2\).

\begin{stw}
If \(G\) is a finite strictly competitive game, resp. quantitatively symmetric game, 
then its mixed extension \(\Delta(G)\) is of the same character.
\end{stw}

\begin{df}
A pair of strategies
\((\sigma_1^{*},\sigma_2^{*})\in\Sigma_{1}\times\Sigma_{2}\) is
\begin{itemize}
\item \textit{Pareto optimum}, \(PO(\Gamma)\), if
\[\neg \exists_{(\sigma_1,\sigma_2)\in\Sigma_{1}\times\Sigma_{2}}
\forall_{i=1,2}\;\; P_{i}(\sigma_1,\sigma_2) >
P_{i}(\sigma_1^{*},\sigma_2^{*}),\]
\item \textit{strong Pareto optimum}, \(SPO(\Gamma)\), if
\[\neg
\exists_{(\sigma_1,\sigma_2)\in\Sigma_{1}\times\Sigma_{2}}\;
\left[\;\forall_{i=1,2}\;\; P_{i}(\sigma_1,\sigma_2) \geq
P_{i}(\sigma_1^{*},\sigma_2^{*}) \;\wedge\; \exists_{i=1,2}\;\;
P_{i}(\sigma_1,\sigma_2) >
P_{i}(\sigma_1^{*},\sigma_2^{*})\;\right],\]
\item \textit{Wald solution}, \(W(\Gamma)\), if
\[\forall_{i=1,2}\;
\sigma_i^{*}\in\arg\max_{\sigma_{i}\in\Sigma_{i}}\;
\min_{\sigma_{-i}\in\Sigma_{-i}}\; P_i((\sigma_i,\sigma_{-i})),\]
\item \textit{Nash equilibrium}, \(NE(\Gamma)\), if
\[\forall_{\sigma_1\in\Sigma_{1},\;\sigma_2\in\Sigma_{2}}
\forall_{i=1,2}\;\; P_{i}((\sigma_{i},\sigma_{-i}^{*})) \leq
P_{i}((\sigma_{i}^{*},\sigma_{-i}^{*})),\]
\item \textit{strict Nash equilibrium}, \(SNE(\Gamma)\), if
\[\forall_{\sigma_1\in\Sigma_{1}\setminus\{\sigma_1^{*}\},
\;\sigma_2\in\Sigma_{2}\setminus\{\sigma_2^{*}\}}
\forall_{i=1,2}\;\; P_{i}((\sigma_{i},\sigma_{-i}^{*})) <
P_{i}((\sigma_{i}^{*},\sigma_{-i}^{*})),\]
\item \textit{semi-strict Nash equilibrium}, \(SSNE(\Gamma)\), if it
is Nash equilibrium and
\[\forall_{\sigma_1\in\Sigma_{1},\;\sigma_2\in\Sigma_{2}}
\forall_{i=1,2}\;\; \left[\;P_{i}((\sigma_{i},\sigma_{-i}^{*})) =
P_{i}((\sigma_{i}^{*},\sigma_{-i}^{*})) \impl
P_{-i}((\sigma_{i},\sigma_{-i}^{*})) =
P_{-i}((\sigma_{i}^{*},\sigma_{-i}^{*}))\;\right],\]
\item \textit{weakly semi-strict Nash equilibrium}, \(WSSNE(\Gamma)\),
if it is Nash equilibrium and
\[\forall_{\sigma_1\in\Sigma_{1},\;\sigma_2\in\Sigma_{2}}
\forall_{i=1,2}\;\; \left[\;P_{i}((\sigma_{i},\sigma_{-i}^{*})) =
P_{i}((\sigma_{i}^{*},\sigma_{-i}^{*})) \impl
P_{-i}((\sigma_{i},\sigma_{-i}^{*})) \geq
P_{-i}((\sigma_{i}^{*},\sigma_{-i}^{*}))\;\right],\]
\item \textit{coupled in wealth improvement}, \(CWI(\Gamma)\), if
\[\forall_{\sigma_1\in\Sigma_{1},\;\sigma_2\in\Sigma_{2}}
\forall_{i=1,2}\;\; \left[\;P_{i}((\sigma_{i},\sigma_{-i}^{*})) \geq
P_{i}((\sigma_{i}^{*},\sigma_{-i}^{*})) \impl
P_{-i}((\sigma_{i},\sigma_{-i}^{*})) \geq
P_{-i}((\sigma_{i}^{*},\sigma_{-i}^{*}))\;\right].\]
\end{itemize}
\end{df}

Remark that \(WSSNE(\Gamma) = NE(\Gamma) \cap CWI(\Gamma)\). 
The last three concepts (SSNE, WSSNE, and CWI) are provided by
ourselves and their role shall be clear in view of further investigations. 
Loosely speaking (weakly) semi-strict equilibria retain most important features 
of strict equilibria, especially those associated with strategic uncertainty 
as shown in Example~\ref{HighThreat}. Remark also that the (weakly) 
semi-strict equilibrium is a concept different from the weakly strict equilibrium 
introduced in \cite{WeakE} and quasi strict equilibrium in \cite{QuasiE}.
Finally the reader should be warned that the Wald solution becomes 
the \textit{maximin solution} if additionally the ``minimax identity" holds
\[\forall_{i=1,2}\;\; P_i(\sigma_1^{*},\sigma_2^{*})=
\max_{\sigma_{i}\in\Sigma_{i}}\;
\min_{\sigma_{-i}\in\Sigma_{-i}}\;
P_i((\sigma_i,\sigma_{-i})).\]

We have the following relations
\[SNE(\Gamma)\varsubsetneq SSNE(\Gamma)\varsubsetneq 
WSSNE(\Gamma)\varsubsetneq NE(\Gamma).\]
The inclusions are immediate from the definitions. That they are strict
illustrates:

\begin{prz}[3-4-5 game]
Let \(S_1=S_2=\{1,2,3,4\}\),
\[G=\left[\begin{array}{cccc}
{[3,3]} & {[0,0]} & {[0,0]} & {[0,0]} \\
{[0,0]} & {[4,4]} & {[0,0]} & {[4,4]} \\
{[0,0]} & {[0,0]} & {[3,3]} & {[5,3]} \\
{[0,0]} & {[4,4]} & {[3,5]} & {[5,5]} \\
\end{array}\right].\]
Then \((4,4)\in NE(G)\setminus WSSNE(G)\), \((3,3)\in
WSSNE(G)\setminus SSNE(G)\), \((2,2)\in SSNE(G)\setminus SNE(G)\),
\((1,1)\in SNE(G)\).
\end{prz}

\begin{stw}
The equilibria of a finite game \(G\) and its mixed extension
\(\Delta(G)\) are related as follows:
\begin{enumerate}
\item \(NE(G)\subset NE(\Delta(G))\),
\item \(SNE(G)\subset SNE(\Delta(G))\),
\item \(SSNE(G)\subset SSNE(\Delta(G))\),
\item \(WSSNE(G)\subset WSSNE(\Delta(G))\).
\end{enumerate}
\end{stw}
\begin{dow}
We only check the last invariance, the rest can be performed
analogously. Let \(((x^{*},y^{*}))\in WSSNE(G)\) and suppose that
some deviation from \(x^{*}\in S_i\) to a mixed strategy \(\pi\in
\Delta(S_i)\) still gives equally good gain for the \(i\)-th
player i.e. \(EW_i((x^{*},y^{*}))=EW_i((\pi,y^{*}))\). First note
that since \(((x^{*},y^{*}))\) is a Nash equilibrium of \(G\), then
\(W_i((x,y^{*}))\leq W_i((x^{*},y^{*}))\) for all \(x\in
\supp\pi\). Now if \(W_i((\sprz{x},y^{*}))< W_i((x^{*},y^{*}))\)
for some \(\sprz{x}\in \supp\pi\), then
\begin{align*}
EW_i((\pi,y^{*})) = \pi(\sprz{x})\cdot W_i((\sprz{x},y^{*})) +
\sum_{\sprz{x}\neq x\in\supp\pi}\; \pi(x)\cdot W_i((x,y^{*})) \\
< \left(\pi(\sprz{x})+\sum_{\sprz{x}\neq
x\in\supp\pi}\;\pi(x)\right) \cdot W_i((x^{*},y^{*})) =
EW_i((x^{*},y^{*})).
\end{align*}
Therefore in fact we have \(W_i((x,y^{*}))=W_i((x^{*},y^{*}))\)
for \(x\in\supp\pi\). Recalling that \(((x^{*},y^{*}))\in
WSSNE(G)\) we obtain that \(W_{-i}((x,y^{*})) \geq
W_{-i}((x^{*},y^{*}))\) for \(x\in\supp\pi\). Finally
\begin{align*}
EW_{-i}((\pi,y^{*})) = \sum_{x\in\supp\pi}\; \pi(x)\cdot
W_{-i}((x,y^{*})) \\
\geq \left(\sum_{x}\; \pi(x)\right) \cdot W_{-i}((x^{*},y^{*})) =
EW_{-i}((x^{*},y^{*})).
\end{align*}
\end{dow}

\begin{tw}\label{antagSSNE}
In the strictly competitive game \(\Gamma\)
all equilibria are semi-strict, \(NE(\Gamma)=SSNE(\Gamma)\).
\end{tw}
\begin{dow}
Let \((\sigma_1^{*},\sigma_2^{*})\in NE(\Gamma)\).
Competitiveness implies that equal payoffs of one player
correspond to equal payoffs of the other one. Therefore
\[P_i((\sigma_i,\sigma_{-i}^{*}))=P_i((\sigma_i^{*},\sigma_{-i}^{*})) \impl
P_{-i}((\sigma_i,\sigma_{-i}^{*}))=P_{-i}((\sigma_i^{*},\sigma_{-i}^{*})),\;
i=1,2,\]
so \((\sigma_1^{*},\sigma_2^{*})\in SSNE(\Gamma)\).
\end{dow}

Recall that the Nash equilibria of the strictly competitive game are precisely maximin solutions.

\section{Lower payoff}

In any game \(\Gamma= (\Sigma_1, \Sigma_2;
P_1,P_2:\Sigma_1\times\Sigma_2\to\R)\) predicted possible
behavior of players may be described via the \textit{response map}
\(R_i: \Sigma_{1}\times\Sigma_{2} \to 2^{\Sigma_{i}}\), \(i=1,2\).
The \textit{best response map}
\(BR_i: \Sigma_{1}\times\Sigma_{2} \to 2^{\Sigma_{i}}\)
defined by
\[BR_{i}((\sigma_{i},\sigma_{-i})) =
\left\{\,\dach{\sigma_{i}}\in\Sigma_{i} \,:\, P_{i}((\dach{\sigma_{i}},\sigma_{-i})) =
\max_{\sigma_{i}'\in\Sigma_{i}}\; P_{i}((\sigma_{i}',\sigma_{-i}))\,\right\}\]
for \(i=1,2\), \(((\sigma_{i},\sigma_{-i})) \in\Sigma_{1}\times\Sigma_{2}\),
is a usual choice, provided that the payoffs are (upper semi-) continuous and 
the strategy spaces are compact.
We would like to investigate other option: not worse response.
The \textit{not worse response map}
\(NWR_i: \Sigma_{1}\times\Sigma_{2} \to 2^{\Sigma_{i}}\) is defined by
\[NWR_{i}((\sigma_{i},\sigma_{-i})) =
\left\{\,\dach{\sigma_{i}}\in\Sigma_{i} \,:\, P_{i}((\dach{\sigma_{i}},\sigma_{-i}))
\geq P_{i}((\sigma_{i},\sigma_{-i}))\,\right\}\]
for \(i=1,2\), \(((\sigma_{i},\sigma_{-i})) \in\Sigma_{1}\times\Sigma_{2}\).

Indeed any player who wants to maximize his payoff would not
change his strategy into a new one, if it leads to lower payoff.
Therefore any response map \(R_i\) should obey the following
restrictions
\[BR_i((\sigma_{i},\sigma_{-i})) \subset
R_i((\sigma_{i},\sigma_{-i})) \subset
NWR_i((\sigma_{i},\sigma_{-i})).\]
Nevertheless one is not forced to play the best response.
We should only incorporate in our calculations the possible answers of the co-player 
to estimate sure gain. This is reflected by the \textit{lower payoff function} 
\(\Pb_i:\Sigma_{1}\times\Sigma_{2}\to\R\),
\[\Pb_i((\sigma_{i},\sigma_{-i})) = 
\inf\; P_{i}((\,\sigma_i, R_{3-i}((\sigma_{-i},\sigma_i))\,)).\]
In the case of our choice \(R_i=NWR_i\):
\[\Pb_i((\sigma_{i},\sigma_{-i})) =
\inf\; \{\,P_{i}((\sigma_i,\dach{\sigma_{-i}})) \,:\,
P_{-i}((\dach{\sigma_{-i}},\sigma_i)) \geq
P_{-i}((\sigma_{-i},\sigma_i)) \,\}.\]

Let us note a simple but useful property
\begin{lem}\label{flatP}
There holds estimation \(\Pb_i\leq P_i\),\, \(i=1,2\).
Moreover, \(\Pb_i(\sigma_1,\sigma_{2})=P_i(\sigma_1,\sigma_{2})\)
for \(i=1,2\) if and only if \((\sigma_1,\sigma_2)\in CWI(\Gamma)\).
\end{lem}

\begin{tw}
Let \(\Sigma_1,\Sigma_2\) be compact spaces.
If \(P_1,P_2:\Sigma_1\times\Sigma_2\to\R\) are continuous,
then \(\Pb_1,\Pb_2:\Sigma_1\times\Sigma_2\to\R\) are lower semicontinuous.
\end{tw}
\begin{dow}
By continuity of \(P_i\) the map
\(NWR_i:\Sigma_{1}\times\Sigma_{2} \to 2^{\Sigma_{i}}\)
has closed graph for \(i=1,2\). Compactness of the domain 
\(\Sigma_1\times\Sigma_2\) implies that the map
\[\Sigma_1\times\Sigma_2\ni ((\sigma_{i},\sigma_{-i})) \mapsto
((\,\sigma_i, NWR_{3-i}((\sigma_{-i},\sigma_i))\,))
\subset\Sigma_1\times\Sigma_2\] is upper semicontinuous with
compact values. Composing it with continuous \(P_i\)
and Hausdorff nonexpansive \(\min: 2^{\R}\to \R\) yields lower
semicontinuity of \(\Pb_i\).
\end{dow}

\begin{stw}\label{antagFlatP}
If \(\Gamma=(\Sigma_1,\Sigma_2; P_1,P_2: \Sigma_1\times
\Sigma_2\to\R)\) is the strictly competitive game,
then for \(\sigma_{i}\in\Sigma_{i}, \sigma_{-i}\in\Sigma_{-i}\)
\[\Pb_{i}((\sigma_i,\sigma_{-i})) =
\inf_{\sigma_{-i}'\in\Sigma_{-i}} P_i((\sigma_i,\sigma_{-i}')).\]
\end{stw}
\begin{dow}
Let \(i=1,2\), \(\sigma_{i},\sigma_{i}'\in\Sigma_{i}\),
\(\sigma_{-i},\sigma_{-i}'\in\Sigma_{-i}\).
Observe that
\[P_{-i}(\sigma_1',\sigma_2')\geq P_{-i}(\sigma_1,\sigma_2) \dstrz
P_{i}(\sigma_1',\sigma_2')\leq P_{i}(\sigma_1,\sigma_2).\]
Hence
\begin{align*}
\Pb_{i}((\sigma_i,\sigma_{-i})) =
\inf \left\{P_i((\sigma_i,\sigma_{-i}))\,:\,
P_{-i}((\sigma_i,\sigma_{-i}'))\geq P_{-i}((\sigma_i,\sigma_{-i}))\right\} = \\
\inf \left\{P_i((\sigma_i,\sigma_{-i}))\,:\,
P_{i}((\sigma_i,\sigma_{-i}'))\leq P_{i}((\sigma_i,\sigma_{-i}))\right\} = 
\inf_{\sigma_{-i}'\in\Sigma_{-i}} P_i((\sigma_i,\sigma_{-i}')).
\end{align*}
\end{dow}

Roughly speaking in the competitive game the lower payoff of the player 
depends only upon his own strategy.

\section{M-equilibrium}\label{sec:m-equilibrium}

We associate with \(\Gamma= (\Sigma_1, \Sigma_2;
P_1,P_2:\Sigma_1\times\Sigma_2\to\R)\) the \textit{flat-game}
\(\Gamma^{\flat}= (\Sigma_1, \Sigma_2;
\Pb_1,\Pb_2:\Sigma_1\times\Sigma_2\to\R)\).

\begin{df}
An \textit{m-equilibrium} of the game \(\Gamma\) is the Nash
equilibrium of \(\Gamma^{\flat}\),
\(ME(\Gamma)=NE(\Gamma^{\flat})\).
\end{df}

The Nash and m-equilibria are not related in a straightforward way.

\begin{prz}\label{MEvsNE}
Let \(S_1=S_2=\{1,2,3\}\),
\[G=\left[\begin{array}{ccc}
{[1,4]} & {[0,0]} & {[4,4]} \\
{[0,0]} & {[3,3]} & {[5,3]} \\
{[4,4]} & {[3,5]} & {[5,3]} \\
\end{array}\right].\]
Then \((2,2)\in ME(G)\cap WSSNE(G)\), \((2,3)\in ME(G) \cap
NE(G)\setminus WSSNE(G)\), \((3,1)\in ME(G)\setminus NE(G)\),
\((3,3)\in NE(G)\setminus ME(G)\) with strongly Pareto optimal
last pair.
\end{prz}

Any decision rule which uses lower payoff estimations (stick
to an m-equilibrium in our case) is resistant on iterated
elimination of dominated strategies. Formally

\begin{stw}
Let the \(i\)-th player, \(i\in\{1,2\}\), expect at least
\(v_i=\Pb_i((\sigma_{i},\sigma_{-i}))\) from playing
\(((\sigma_{i},\sigma_{-i}))\) in \(\Gamma\). Suppose that the partner
of \(i\) changes his strategy \(\sigma_{-i}\) into \(\sigma_{-i}'\)
according to possibly higher payoff
\(P_{-i}((\sigma_{-i}',\sigma_i)) \geq P_{-i}((\sigma_{-i},\sigma_i))\). 
Then the \(i\)-th player is still satisfied, because 
\(P_{i}((\sigma_i,\sigma_{-i}')) \geq v_i\).
\end{stw}

This obvious assertion (restatement of the definition of the lower payoff) 
explains why the player does not need to reject his dominated strategies.

To understand possible quirks of being content with warranted
lower payoff one should consider two classic zero-sum games.

\begin{prz}[Hide a coin]\label{hidecoin}
Let \(S_1=S_2=\{1,2\}\),
\[G=\left[\begin{array}{cc}
{[-10,10]} & {[15,-15]} \\
{[15,-15]} & {[-20,20]}  \\
\end{array}\right].\]
Then \(NE(G)=\emptyset\), although
\(ME(G)=NE(G^{\flat})=\{1\}\times S_2\),
\[G^{\flat}=\left[\begin{array}{cc}
{[-10,-15]} & {[-10,-15]} \\
{[-20,-15]} & {[-20,-15]}  \\
\end{array}\right].\]
The first player cannot ensure payoff higher than \(-10\), the
second player cannot ensure payoff higher than \(-15\).
\end{prz}

\begin{prz}[Matching pennies]\label{pennies}
Let \(S_1=S_2=\{1,2\}\),
\[G=\left[\begin{array}{cc}
{[1,-1]} & {[-1,1]} \\
{[-1,1]} & {[1,-1]}  \\
\end{array}\right].\]
Then \(NE(G)=\emptyset\), although
\(ME(G)=NE(G^{\flat})=S_1\times S_2\),
\[G^{\flat}=\left[\begin{array}{cc}
{[-1,-1]} & {[-1,-1]} \\
{[-1,-1]} & {[-1,-1]}  \\
\end{array}\right].\]
None of the players can ensure payoff higher than \(-1\).

On the other hand \(NE(\Delta(G)) = 
\{\left(\frac{1}{2}\delta_1+\frac{1}{2}\delta_2, 
\frac{1}{2}\delta_1+\frac{1}{2}\delta_2\right)\}\)
and this constitutes the main argument for mixed strategies
if we view strategic interaction as did von Neumann: trying to
outmanoeuvre other participants.
\end{prz}

The key to resolve inconsistency of m-equilibrium with the hiding
player's choice policy behind a mixed strategy is to recognize that
the m-equilibrium concentrates on the question \emph{what can be
warranted in one-shot game} rather than 
\emph{what can be gambled during repeated play}. This seems
paradoxical, but one also needs to take into account injurious
though rational player as during analysis in
Example~\ref{HighThreat} (comp. comment from
p.\pageref{exploitation} before Theorem~\ref{WSSNEisME}).

We investigate further properties of flat-games and m-equilibria.

\begin{stw}
Lower value of a game does not change when substitute payoffs
with lower payoffs:
\[\sup_{\sigma_{i}\in\Sigma_{i}}\;\;\inf_{\sigma_{-i}\in\Sigma_{-i}}\;\;
P_i((\sigma_{i},\sigma_{-i}))=
\sup_{\sigma_{i}\in\Sigma_{i}}\;\;\inf_{\sigma_{-i}\in\Sigma_{-i}}\;\;
\Pb_i((\sigma_{i},\sigma_{-i})).\]
\end{stw}
\begin{dow}
Fix \(\sigma_i\in\Sigma_i\), \(\sigma_{-i}\in\Sigma_{-i}\), \(i=1,2\).
By the definition of the lower payoff and Lemma~\ref{flatP}:
\[\inf_{\sigma_{-i}\in\Sigma_{-i}}\;\;P_i((\sigma_{i},\sigma_{-i})) \leq
\Pb_i((\sigma_{i},\sigma_{-i})) \leq P_i((\sigma_{i},\sigma_{-i})).\]
Thus
\begin{equation}
\inf_{\sigma_{-i}\in\Sigma_{-i}}\;\;P_i((\sigma_{i},\sigma_{-i})) =
\inf_{\sigma_{-i}\in\Sigma_{-i}}\;\;\Pb_i((\sigma_{i},\sigma_{-i})).
\end{equation}
\end{dow}

M-equilibrium is a strategic concept -- it depends upon mutual
preferences of players.

\begin{stw}
Let \(\Gamma=(\Sigma_1,\Sigma_2; P_1,P_2: \Sigma_1\times
\Sigma_2\to\R)\) be a game and \(\varphi_i: \R\to\R\)
be strictly increasing right continuous functions, \(i=1,2\).
Then the game
\(\widetilde{\Gamma}=(\Sigma_1,\Sigma_2;
\widetilde{P_1},\widetilde{P_2}:\Sigma_1\times\Sigma_2\to\R)\)
transformed from \(\Gamma\) via the formula
\(\widetilde{P_i}=\varphi_{i}\circ P_i\) admits the same
m-equilibria as the original game \(\Gamma\); symbolically
\(ME(\widetilde{\Gamma}) = ME(\Gamma)\).
\end{stw}
\begin{dow}
Observe that \(\varphi_{i}(\inf Z) = \inf \varphi_{i}(Z)\) for \(Z\subset\R\), \(i=1,2\).
Then a direct calculation shows that
\(\widetilde{P_i}^{\flat} = \varphi_{i}\circ \Pb_i\)
whence the conclusion follows.
\end{dow}

We postpone a more technically subtle discussion of the above property
to the Appendix.

Weakly semi-strict Nash equilibria are those equilibria which
survive ``flattenization" of the game. Due to a one-sided {exploitation of
player's trust}\label{exploitation} the other player might change
his strategy without loss of his payoff just to lower the payoff
of his partner, which explains why not all Nash equilibria are
m-equilibria (cf. Example~\ref{HighThreat}). Despite possible
complications illustrated by Example~\ref{MEvsNE} a positive
criterion for a Nash equilibrium to be an m-equilibrium provides

\begin{tw}\label{WSSNEisME}
The equilibria of the game \(\Gamma\) and its flat
\(\Gamma^{\flat}\) are related as follows:
\begin{enumerate}
\item \(WSSNE(\Gamma)\subset WSSNE(\Gamma^{\flat})\),
\item \(SSNE(\Gamma)\subset SSNE(\Gamma^{\flat})\),
\item \(SNE(\Gamma)\subset SNE(\Gamma^{\flat})\).
\end{enumerate}
\end{tw}
\begin{dow}
We only check the first inclusion, the rest being analogous.

Let \((\sigma_1^{*},\sigma_2^{*})\in WSSNE(\Gamma)\), \(i=1,2\).
By Lemma~\ref{flatP} for \(\sigma_i\),
\begin{equation}\label{PbPPPb}
\Pb_i((\sigma_i,\sigma_{-i}^{*})) \leq
P_i((\sigma_i,\sigma_{-i}^{*})) \leq
P_i((\sigma_i^{*},\sigma_{-i}^{*})) =
\Pb_i((\sigma_i^{*},\sigma_{-i}^{*})),
\end{equation}
which shows \((\sigma_1^{*},\sigma_2^{*})\in NE(\Gamma^{\flat})\).

Suppose that \(\Pb_i((\sigma_i,\sigma_{-i}^{*})) =
\Pb_i((\sigma_i^{*},\sigma_{-i}^{*}))\). Then from (\ref{PbPPPb})
\(P_i((\sigma_i,\sigma_{-i}^{*}))
=P_i((\sigma_i^{*},\sigma_{-i}^{*}))\). Observe that
\begin{align*}
\Pb_{-i}((\sigma_i,\sigma_{-i}^{*})) = \inf\left\{\,
P_{-i}((\dach{\sigma_i},\sigma_{-i}^{*})) \,:\,
P_{i}((\dach{\sigma_i},\sigma_{-i}^{*})) \geq
P_{i}((\sigma_i,\sigma_{-i}^{*})) =
P_{i}((\sigma_i^{*},\sigma_{-i}^{*})) \,\right\} \\
= \inf\left\{\, P_{-i}((\dach{\sigma_i},\sigma_{-i}^{*})) \,:\,
P_{i}((\dach{\sigma_i},\sigma_{-i}^{*})) =
P_{i}((\sigma_i^{*},\sigma_{-i}^{*})) \,\right\},
\end{align*}
since \((\sigma_1^{*},\sigma_2^{*})\) stays in equilibrium. Now
any \(\dach{\sigma_i}\) with
\(P_{i}((\dach{\sigma_i},\sigma_{-i}^{*})) =
P_{i}((\sigma_i^{*},\sigma_{-i}^{*}))\) gives
\(P_{-i}((\dach{\sigma_i},\sigma_{-i}^{*})) \geq
P_{-i}((\sigma_i^{*},\sigma_{-i}^{*}))\), because
\((\sigma_1^{*},\sigma_2^{*})\) is weakly semi-strict equilibrium.
Hence
\[\Pb_{-i}((\sigma_i,\sigma_{-i}^{*})) \geq
P_{-i}((\sigma_i^{*},\sigma_{-i}^{*})) =
\Pb_{-i}((\sigma_i^{*},\sigma_{-i}^{*})),\]
where the last equality assures Lemma~\ref{flatP}.
\end{dow}

\begin{stw}
If \(\Gamma\) is quantitatively symmetric, then \(\Gamma^{\flat}\) too.
\end{stw}

Neither zero-sum, nor strict competitiveness of the game is preserved under 
``flattenization" procedure as show Examples~\ref{hidecoin} and~\ref{pennies}.

\section{Motivating examples}\label{sec:motivexamples}

We bring to the readers attention three classic games where the m-equilibrium turns out
to be a nontrivial concept. 

\begin{prz}[Traveler's dilemma -- continuation]
Let \(G\) be as in Example \ref{traveler}. We have
\(NE(G)=SNE(G)=\{(2,2)\}\) and
\({\Pb}_1(x,y) = {\Pb}_2(y,x)= \min(x,y)-4+2\cdot\sign(x-y)\) 
for \(x,y\in \{2,3,\ldots,100\}\).
Hence \(G^{\flat}\) admits two equilibria, so that
\(ME(G)=\{(2,2),(100,100)\}\). The outcome \((100,100)\) 
was often proposed by people (cf. \cite{SciAm}) as a reasonable Pareto optimal solution 
regardless of a possible treacherous behavior of the co-player. 
(Interestingly, \(G^{\flat\flat}\) possesses three equilibria, which shows 
that \(NE(G^{\flat\flat})\neq NE(G^{\flat})\) in general).
\end{prz}

\begin{prz}[Cournot duopoly; \cite{Oligopolies,Strategy}]\label{CournotDuopoly}
Let \(\Gamma= (\Sigma_1, \Sigma_2; P_1,P_2:\Sigma_1\times\Sigma_2\to\R)\),
\(S_1=S_2=[0,L]\), \(L>0\), 
\(P_1(x,y)=P_2(y,x) = x\cdot \left(L-(x+y)\right)\) for \(x,y\in [0,L]\) .
Then \(NE(\Gamma)= SNE(\Gamma) = \left(\frac{L}{3},\frac{L}{3} \right)\).

Using elementary methods (cf. \cite{Strategy}) we find that 
\({\Pb}_1(x,y)= x\cdot \min(y, L-(x+y))\) and
\[
ME(\Gamma) = \{(x,L-2x) \,:\, 0\leq x \leq {L}/{3}\} \cup
\left\{ \left(x, \frac{L-x}{2}\right) \,:\, {L}/{3}\leq x \leq L\right\}.
\]

Note that from the cartel point of view, a Pareto dominating TU-solution
\(({L}/{4},{L}/{4})\) would be superior. Unfortunately this ``natural" solution 
is not an m-equilibrium. Nevertheless the joint payoff
\(P_1+P_2\) is maximized at two boundary m-equilibria: \((0,L)\) and
\((L,0)\). This suggests that under Cournot duopoly pricing it is profitable
for firms to choose an active monopolist and the other firm rest with no production. 
Switching the role of monopolist between firms could become a strategy
(in repeated game) for hidden transfer of utility despite 
the payoff in game was assumed to be NTU.
\end{prz}

\begin{prz}[Puu duopoly; \cite{Oligopolies,PuuOligopolies}]
Let \(\Gamma= (\Sigma_1, \Sigma_2; P_1,P_2:\Sigma_1\times\Sigma_2\to\R)\),
\(S_1=S_2=[0,L]\), \(L>1\), 
\[
P_1(x,y)=P_2(y,x) = \left(\frac{L}{x+y}-1\right) \cdot x
\]
for \(x,y\in [0,L]\) with convention that \(P_i(0,0)=0\), \(i=1,2\).
Then \(NE(\Gamma) = \left(\frac{L}{4},\frac{L}{4} \right)\).

By elementary (though a bit cumbersome) calculations
\[
{\Pb}_1(x,y) = \left(\frac{L}{x+{y}^{\sharp}}-1\right) \cdot x
\]
for \((x,y)\neq (0,0)\), where
\({y}^{\sharp} = \max\left(y, \left(\frac{L}{x+y}-1\right) \cdot x\right)\).
Hence \({\Pb}_1(x,y)={\Pb}_2(y,x) = \min(y,P_1(x,y))\) for all \(x,y\in[0,L]\).

Puu duopoly enjoys a rich set of m-equilibria. 
Denote by \(L_{*}\approx 3.0796\) the unique positive root
of the polynomial \(1+4\,L+6\,L^2+4\,L^3+L^4-L^5\) and put
\[
E=\left\{\begin{array}{ll}
\{(\sqrt{L},\sqrt{L}\cdot(\sqrt[4]{L}-1)), (\sqrt{L}\cdot(\sqrt[4]{L}-1),\sqrt{L}) \}, & 
\mbox{ when } L=L_{*},\\
{\emptyset}, & \mbox{ otherwise}.\\
\end{array}\right.
\]
Further, denote 
\[
N=\left\{\begin{array}{ll}
\{\left({L}/{4},{L}/{4}\right) \}, & 
\mbox{ when } L>16,\\
{\emptyset}, & \mbox{ otherwise}.\\
\end{array}\right.
\]
We have
\[
ME(\Gamma)= \bigcup_{x\in[0,\sqrt{L}]} \{x\}\times [0,\sqrt{L}-x] \,\cup E \cup N.
\]

The extraordinary pair of m-equilibria at \(L=L_{*}\) is an unexpected phenomenon.
(It seems to be a noneconomic artifact bond to the formal model).
That Nash equilibria of \(\Gamma\) need not be m-equilibria unless \(L\) is sufficiently 
large, is an effect of weakness of Nash equilibrium: when taking into account 
the security of payoff, a treacherous partner can switch precomitted (during
cheap talk) strategy to a strategy indifferent for him but harmful for his co-player. 
Formally, \(({L}/{4},{L}/{4})\) is not a
(weakly semi-) strict Nash equilibrium for small \(L\).

Finally observe that for \(L<4\) the m-equilibrium \(({\sqrt{L}}/{2},{\sqrt{L}}/{2})\)
Pareto dominates the Nash equilibrium \(({L}/{4},{L}/{4})\). One can stipulate that
such m-equilibria might ``explain" cartels in a game theoretic way without 
a recourse to exterior (outside game) constructs.
\end{prz}

There is no doubt that the traveler's dilemma was the driving force
of our research. Note that m-equilibria do not bring anything new 
to the (in)famous prisoner's dilemma (PD). This confirms that 
the traveler's dilemma is not merely an extension of PD -- it is something 
qualitatively different. On the other hand simultaneous simplicity and nontriviality 
of PD shows that having a good solution concept does not negate the reason 
to perform the play at all: knowing consequences is not freeing us 
from making decisions.

\section{Existence of m-equilibrium}

We know from Theorem~\ref{WSSNEisME} that the class of games which possess 
at least one m-equilibrium is quite large. Unfortunately we do not know whether
m-equilibria exist under suitaby mild assumptions in general. 
Nevertheless competitive games admit pure m-equilibria.

\begin{tw}
If \(\Gamma=(\Sigma_1,\Sigma_2; P_1,P_2: \Sigma_1\times\Sigma_2\to\R)\)
is the strictly competitive game
with compact metrizable strategy spaces \(\Sigma_i\) and
continuous payoffs \(P_i\), \(i=1,2\),
then \(ME(\Gamma)\neq\emptyset\).
Namely, Wald solutions rest in m-equilibrium.
\end{tw}
\begin{dow}
Observe that for \(\sigma_i,\sigma_i'\in\Sigma_i\), \(i=1,2\)
\[d_H\left(\,P_i((\sigma_i,\Sigma_{-i})),P_i((\sigma_i',\Sigma_{-i}))\,\right)
\leq \sup_{\sigma_{-i}\in\Sigma_{-i}}
\left|P_i((\sigma_i,\sigma_{-i})) - P_i((\sigma_i',\sigma_{-i}))\right|,\]
where \(d_H\) stands for the Hausdorff distance in \(2^{\R}\).
Since \(P_i\) are uniformly continuous (as continuous on the compactum),
we know that \(\Psi_i:\Sigma_i\to 2^{\R}\),
\(\Psi_i(\sigma_i) = P_i((\sigma_i,\Sigma_{-i}))\) for \(\sigma_i\in\Sigma_i\), 
are Hausdorff continuous with compact values. The Hausdorff nonexpansiveness 
of \(\min:2^{\R}\to\R\) yields then the continuity of the map
\[\Sigma_i\ni \sigma_i\mapsto \min_{\sigma_{-i}\in\Sigma_{-i}} 
P_i((\sigma_i,\sigma_{-i})) = \min \Psi_i(\sigma_i).\]
This shows that we can define
\[\sigma_i^{*} \in \arg\max_{\sigma_i\in\Sigma_i}\; 
\min_{\sigma_{-i}\in\Sigma_{-i}}\, P_i((\sigma_i,\sigma_{-i}))\]
for \(i=1,2\).
So \((\sigma_1^{*},\sigma_2^{*})\in W(\Gamma)\neq\emptyset\).

Further by Proposition~\ref{antagFlatP}
\begin{align*}
\max_{\sigma_i\in\Sigma_i}\; \min_{\sigma_{-i}\in\Sigma_{-i}}\, 
P_i((\sigma_i,\sigma_{-i})) =
\max_{\sigma_i\in\Sigma_i}\; \Pb_i((\sigma_i,\sigma_{-i})) \\
= \max_{\sigma_i\in\Sigma_i}\; \Pb_i((\sigma_i,\sigma_{-i}^{*}))
= \Pb_i((\sigma_{i}^{*},\sigma_{-i}^{*}))
\geq \Pb_i((\sigma_i,\sigma_{-i}^{*}))
\end{align*}
for any \(\sigma_i\in\Sigma_i\), \(\sigma_{-i}\in\Sigma_{-i}\).
Therefore
\(W(\Gamma)\subset NE(\Gamma^{\flat}) = ME(\Gamma)\).
\end{dow}

Unfortunately the results established so far in the literature 
(cf. \cite{Bagh,Reny,Scalzo}) which are concerned with the existence 
of equilibria in games with discontinuous payoff functions 
do not seem to be applicable for the kind of problems considered here.

\section{Mixed strategies and risk}

The reason to calculate lower payoffs is establishing sure gains. Therefore one might question 
the use of expected payoffs to evaluate gains. 
We single out this phenomenon in the case of zero sum game.

\begin{prz}[extended matching pennies]
Let \(S_1=S_2=\{1,2,3\}\),
\[G=\left[\begin{array}{ccc}
{[-1,1]} & {[1,-1]} & {[0,0]} \\
{[1,-1]} & {[-1,1]} & {[0,0]} \\
{[0,0]} & {[0,0]} & {[0,0]} \\
\end{array}\right].\]
Then \(NE(\Delta(G)) = \{(\delta_3,\delta_3)\),
\(\left(\frac{1}{2}\delta_1+\frac{1}{2}\delta_2, \frac{1}{2}\delta_1+\frac{1}{2}\delta_2\right)\), 
\(\left(\frac{1}{2}\delta_1+\frac{1}{2}\delta_2,\delta_3\right)\), 
\(\left(\delta_3,\frac{1}{2}\delta_1+\frac{1}{2}\delta_2\right)\}\).
Although all equilibria yield the same expected payoff, they differ significantly 
from the point of view of the risk. Namely the variance in payoff is nonzero unless 
both players use pure strategies (standard property of random variables). 
This has consequence for risk averse players usually not considered in the 
classic von Neumann's minimax theory.
\end{prz}

Let \(S=\{x_1,x_2,x_3,x_4,\ldots\}\) be the set of prizes with the associated utility function 
\(U:S\to \R\), such that \(U(x_1)<U(x_3)<U(x_2)\).
Risk neutral players calculate gain for the lottery \((S,\pi)\), \(\pi\in\Delta(S)\), via the expected utility
\[EU(\pi)=\sum_{x\in S} \pi(x)\cdot U(x).\]
Hence they are indifferent in the choice between two lotteries 
\((\{x_1,x_2\},\rho)\) and \(\{x_3\}\) as long as \(EU(\rho)=U(x_3)\).

However loss averse players would rather calculate the minimal gain 
\[E^{\min}U(\pi)= \min_{x\in\supp\pi} U(x)\]
for the lottery \((S,\pi)\).
Then \(E^{\min}U(\rho)<U(x_3)\) and \(\{x_3\}\) is preferred over \((\{x_1,x_2\},\rho)\) 
whenever \(x_1\in\supp\rho\).
Take into account another lottery \((\{x_1,x_2\},\rho')\) such that \(x_1\in\supp\rho'\).
Then \(E^{\min}U(\rho')=E^{\min}U(\rho)\), so \(\rho\) and \(\rho'\) seem equally good.
Still loss averse players might evaluate which of the given two lotteries with the same minimal gain 
has higher expected gain (as secondary criterion for preferences), \(EU(\rho)\) or \(EU(\rho')\)?

Concerning mixed strategies one should also be aware that the probability distribution might be
also interpreted deterministically as a set of weights describing ``fair" allocation of welfare/payoff
induced by the choice of strategies. We discuss related questions in the next Section.

\section{Equilibrium selection}\label{sec:equilibriumselection}

The concept of m-equilibrium takes loss aversion and correlated decision into serious consideration.
It demands communication and sure gains to be estimated. However it is 
\emph{not} correlated equilibrium of Aumann. It also accounts for losses on the more 
basic level than the Harsanyi-Selten risk dominance selection criterion. 
Nevertheless m-equilibria (being Nash equilibria of the game with flattened payoffs) suffer 
the same curse of nonuniqueness (of payoff) as other notions of solution designed 
for non zero sum games.

To avoid complicated matters of the formal definitions of  communication (pre-play) we simply say
that the \textit{players can communicate} to establish the final decision in a game 
\(\Gamma=(\Sigma_1,\Sigma_2; P_1,P_2: \Sigma_1\times\Sigma_2\to\R)\), provided there 
exists a ``communication channel" \(\mathcal{C}: \Sigma_1\times\Sigma_2 \to ME(\Gamma)\), 
where \(\mathcal{C}\) is a random variable distributed on the set of m-equilibria according to
probability \(\alpha\in \Delta[ME(\Gamma)]\). Vector \(\alpha\) will be interpreted further
also as the set of weights of welfare allocation among m-equilibria. 

Although communication restores Pareto-efficient equilibrium selection in 
Example \ref{Coordination} and the stag hunt game, we will face 
classical coordination dilemma posed by the battle of the sexes game.
In presence of equal power and credibility players, the coordination dilemma
is often resolved via fair allocation rule: ``once for me, once for you". We believe
that cooperative social choice among various equilibria is the appropriate answer
to equilibrium selection in both, one-shot and repeated games. Together with
a social rule providing the allocation vector \(\alpha\in\Delta[ME(\Gamma)]\), some
stochastic tie-breaking rules are indispansable. (During repeated play the variance
of payoff outcomes arises as another problem. Alternate choice of equilibria minimizes
this variance).

\begin{prz}[Battle of the sexes]
Let \(S_1=S_2=\{1,2\}\),
\[G=\left[\begin{array}{cc}
{[3,2]} & {[0,0]} \\
{[0,0]} & {[2,3]}  \\
\end{array}\right].\]
Then \(NE(G)=ME(G) = \{(1,1), (2,2)\}\); 
\(\alpha = \frac{1}{2}\cdot \delta_{(1,1)} + \frac{1}{2}\cdot \delta_{(2,2)}\).
The only way to get rid of the question ``who's equilibrium played first" is 
to apply randomization device according to distribution given by \(\alpha\).
This is an instance of Szaniawski's probabilistically equal choice principle (\cite{Lissowski}).
In one-shot games stochastic mechanism for choosing the player who selects preferred
equilibrium to be played seems very reasonable also according to Laplace's criterion of
insufficient reason.
\end{prz}

Once players receive the recommended equilibrium after pre-play phase, they form their 
beliefs and strategic properties of m-equilibrium warrant the appropriate payoff levels 
regardless of whether one of the players tries to exploit this information. 
Roughly speaking, to cut-off the inductive race to the bottom, 
it is enough that at least one player is fair. 
This unavoidably leads to the problem of reputation.

\section{Final comments}

The following problems are very important for the discussion of the relevance of the concept 
of m-equilibrium:
\begin{enumerate}
\item Do (pure strategy) m-equilibria always exist under reasonable assumptions about 
payoff functions?
\item  What other than traveler's dilemma games admit ``intuitively superior" m-equilibria 
impossible within standardly interpreted Nash framework?
\item How to cope with risk and welfare allocation? Does there exist any clear risk dominance 
criterion? (Cf. \cite{Harsanyi}).
\end{enumerate}

To prove a general existence theorem on m-equilibria definitely one cannot use continuity of 
lower payoffs, but some assumptions about payoffs and strategy sets are indispensable.

\begin{prz}
Let \(\Gamma=(\Sigma_1,\Sigma_2; P_1,P_2: \Sigma_1\times\Sigma_2\to\R)\), 
\(x,y\in\Sigma_1=\Sigma_2=[0,\infty)\) and \(L\geq 2C>0\). We define
\(P_1(x,y)=P_2(y,x)= \frac{Lx}{x+y} - \frac{C}{x}\), when \(x>0\) and 
\(P_1(0,y)=0\) otherwise. We assume here (unlike\cite{PuuOligopolies, Oligopolies}) 
that the total cost of production decreases \(C/x \searrow 0\) as 
the production of the firm increases \(x\nearrow \infty\).
One can think about this opportunity as the effect of scale (globalization). It turns out that under
our assumption of diminishing cost, the Puu duopoly behaves qualitatively in a similar fashion to
that observed in the competition of ``giants": each player has an incentive to grow production 
for overtaking the market; in practice we expect the mirroring behavior of firms, 
because it warrants maximal payoff share (according to TU value).

Interestingly \({\Pb}_1(x,y)={\Pb}_2(y,x) = -\,\frac{C}{x}\), which reflects an old truth that 
in business one might bear the cost of production without any profit (``fall of a giant"). 
Consequently \(ME(\Gamma)= \{(0,0)\}\). If \(\Sigma_1=\Sigma_2=(0,\infty)\), then 
\(\Gamma\) has no m-equilibrium. A reasonable workaround could be then to allow for 
an epsilon-m-equilibrium (produce as little as possible).
\end{prz}

A sky-rocketing competition in the above Example tells us that the dynamic view of games
is necessary when the game is played more than once.

We do not consider multiplayer games in this article because we believe that only 
good understanding of two player games can give rise for reasonable extensions of static duel 
games to the situation of multiple interacting agents. We are aware of specific ``phase transitions"
and emergent effects when passing from the case of two players to the case of multiple players.

The adaptation of the notion of m-equilibrium for multiplayer games should be done carefully.
Let  \(\Gamma= (\Sigma_1, \ldots, \Sigma_N; 
P_1,{\ldots},P_N:\Sigma_1\times\ldots\times\Sigma_N\to\R)\) 
be a game with \(N\) players.
Since the player can only be sure of his own declaration and the communicated decisions of others
might be changed, the following definition of the lower payoff seems to be suitable in this sort
of situation:
\begin{gather*}
\Pb_i(\sigma_{1},{\ldots},\sigma_{N}) = \inf\; 
\{\,P_{i}({\varsigma}^{J}) \\
\,:\, \exists_{J\subset \{1,{\ldots},N\}}\;\forall_{j\in J}\;\; 
P_{j}({\varsigma}^{J}) 
\geq P_{j}(\sigma_{1},{\ldots},\sigma_{N}) \,\}
\end{gather*}
for \((\sigma_{1},{\ldots},\sigma_{N})\), 
\({\varsigma}^{J}\)
\(\in\Sigma_1\times\ldots\times\Sigma_N\),
where \({\varsigma}^{J}_i = \sigma_i\) when \(i\not\in J\), i.e., \({\varsigma}^{J}\)
may differ from \((\sigma_{1},{\ldots},\sigma_{N})\) for players \(i\) contributing
to a virtual coalition \(J\).

Some other ideas aiming to resolve the dominated strategies conundrum in traveler's dilemma 
were reported in \cite{HalpernPass}. Another concept of solution suitable for traveler's dilemma
(and accounting for the lack of communication unlike in the present article) is Hofstadter's 
superrationality which can be argued within bayesian framework, e.g., \cite{QuasiMagical}.  
However our intention was to dispose off as much probability as possible.

The cryptic term ``m-equilibrium" was thought off by the author in accordance 
with the notion of meta-stable equilibrium appearing among others in chemistry and physics; 
that is an extraordinary equilibrium (or higher state) possible only under 
very specific conditions.

\section*{Appendix: Isomorphism of games}

We say that a function \(\varphi: \R\supset Z \to \R\) is
\begin{itemize}
\item \textit{strictly inf-increasing}, if for nonempty
\(U_1,U_2\subset Z\)
\[\inf U_1<\inf U_2 \Rightarrow \inf\varphi(U_1) < \inf\varphi(U_2),\]
\item \textit{inf-continuous}, if for every nonempty \(U\subset Z\) 
such that \(\inf U\in Z\) holds \(\varphi(\inf U) = \inf\varphi(U)\),
\item \textit{right continuous}, if for every \(z_0\in Z\) and every
(w.l.o.g. decreasing) sequence \(z_n\in Z\), \(z_0\leq z_n\to z_0\) holds
\(\varphi(z_n)\to\varphi(z_0)\).
\end{itemize}

\begin{stw}
Let \(\varphi: \R\supset Z\to\R\).
\begin{enumerate}
\item If \(\varphi\) is strictly inf-increasing, then it is strictly increasing.
\item If \(\varphi\) is (not necessarily strictly) increasing, then
it is inf-continuous if and only if it is right continuous.
\item If \(Z=\R\) and \(\varphi\) is strictly increasing inf-continuous, then it is strictly inf-increasing.
\end{enumerate}
\end{stw}

We warn that infima are taken in the whole \(\R\), not in the ordered subset \(Z\subset\R\). 
That the notion of strictly inf-increasing function is essentially stronger 
than strictly increasing function illustrates

\begin{prz}
Let \(Z=\{0\}\cup (1,\infty)\subset\R\), \(\varphi: Z\to\R\),
\(\varphi(z)=\max\{z-1,0\}\) for \(z\in Z\).
Although \(\varphi\) is strictly increasing inf-continuous function
it is not strictly inf-increasing.
\end{prz}

\begin{stw}
Let \(\Gamma=(\Sigma_1,\Sigma_2; P_1,P_2: \Sigma_1\times
\Sigma_2\to\R)\) be a game and \(\varphi_i: P_i(\Sigma_1\times
\Sigma_2)\to\R\) be order-preserving maps, \(i=1,2\), i.e.
\[\forall_{u,v\in P_i(\Sigma_1\times \Sigma_2)}\;\;
u<v \impl \varphi_{i}(u)<\varphi_{i}(v).\]
Then the game
\(\widetilde{\Gamma}=(\Sigma_1,\Sigma_2;
\widetilde{P_1},\widetilde{P_2}:\Sigma_1\times\Sigma_2\to\R)\)
transformed from \(\Gamma\) via the formula
\(\widetilde{P_i}=\varphi_{i}\circ P_i\) admits the same
m-equilibria as the original game \(\Gamma\); symbolically
\(ME(\widetilde{\Gamma}) = ME(\Gamma)\).
\end{stw}
\begin{dow}
Direct calculation shows that \(\widetilde{P_i}^{\flat} =
\varphi_{i}\circ \Pb_i\) whence the conclusion follows.
[Needed \(\varphi(\inf Z)= \inf \varphi(Z)\)].
\end{dow}

An analysis of ``equivalent" prisoner's dilemmas shows that isomorphic
games may have nonequivalent risk structure. Therefore an appropriate
concept of isomorphism of normal form games is no less controversial
than the choice of satisfactory definition of the solution of a game or 
the equilibrium selection problem.

\section*{Acknowledgement}

The author's research was ignited in 2008 by S{\l}awomir Plaskacz
(differential inclusions, Hamilton-Jacobi equation in control and optimization, 
differential games). We had a lot of vigorous discussions. 

A criticism of the earlier concepts proposed by the author (Nash-von Neumann
cooperative solution, correlated Pareto optimum, retaliatory safe optimum)
led to the concept of m-equilibrium.
I would like to thank all participants of the three seminars where, around 2009, 
I referred those unsatisfactory concepts: 
Seminar of the Chair of Nonlinear Mathematical Analysis and Topology
at the Nicolaus Copernicus University (Wojciech Kryszewski's group), 
Seminar of the Game and Decision Theory Group at the Polish Academy of Sciences
(Andrzej Wieczorek's group) and Seminar of the Chair of Mathematical Economics 
at the Poznan University of Economics (Emil Panek's group).
 
Almost all of this work has been done by the author during 2009-2010 in the
Faculty of Mathematics and Computer Science at the Nicolaus Copernicus University
(Toru\'{n}, Poland).

It would be really hard to grasp the current state of the research in game theory,
if not \textit{books.google} and various free resources provided by the experts 
in the subject.

\end{document}